# One-third magnetization plateau in a spin-1 kagome magnet BaNi$_3$(AsO$_4$)$_2$(OH)$_2$

Yuya Haraguchi[1,†], Jun-ichi Yamaura[2], Akira Matsuo[2], Koichi Kindo[2], and Hiroko Aruga Katori[1]

[1]Department of Applied Physics and Chemical Engineering, Tokyo University of Agriculture and Technology, Koganei, Tokyo 184-8588, Japan
[2]The Institute for Solid State Physics, The University of Tokyo, Kashiwa, Chiba 277-8581, Japan
[†]Corresponding author: chiyuya3@go.tuat.ac.jp

We investigate the structural and magnetic properties of BaNi$_3$(AsO$_4$)$_2$(OH)$_2$, focusing on its spin-1 kagome lattice and the intricate coexistence of ferromagnetic and antiferromagnetic interactions. Powder x-ray diffraction analysis confirms a highly crystalline trigonal structure. Detailed Rietveld refinement identifies a single crystallographic Ni site, indicative of a perfect kagome lattice. Magnetic susceptibility measurements suggest predominantly ferromagnetic interactions with an effective magnetic moment consistent with Ni$^{2+}$ spins, yet the system undergoes antiferromagnetic ordering at a Néel temperature of 5.8 K. Isothermal magnetization measurements reveal a series of metamagnetic transitions culminating in a plateau-like phase near one-third of the total saturation magnetization. Analysis of the phase boundaries shows that the antiferromagnetic phase supports a substantial net moment in each kagome layer, comparable to that of the one-third plateau. This observation challenges the conventional model—where a 120° ground state transitions to an up-up-down configuration—commonly assumed for kagome antiferromagnets. Instead, our findings indicate that both the zero-field ground state and the field-induced phases exhibit in-plane ferrimagnetic spin arrangements on the kagome lattice, with the metamagnetic transition corresponding to a shift from layer-by-layer antiferromagnetically aligned net moments to ferromagnetically aligned ones. This configuration is stabilized by bond frustration, a network of competing interactions that can favor both ferromagnetic and antiferromagnetic couplings, highlighting the essential role of frustration in governing the low-temperature magnetic behavior of spin-1 kagome systems. By moving beyond the canonical 120° paradigm, these results expand the landscape of unconventional magnetic phases and emphasize the intricate interplay between geometry and frustration in kagome lattices.

## I. Introduction

Highly frustrated magnetic systems have long been a focal point in condensed matter physics, providing a unique platform to explore novel quantum phenomena arising from extensive degeneracy [1-5]. One of the most captivating and elusive structures in this context is the kagome lattice, derived from a traditional Japanese woven basket pattern. The kagome lattice consists of corner-sharing triangles, leading to a rich variety of quantum phases due to its significant geometric frustration.

Despite extensive research, the exact solution to the kagome-lattice Heisenberg antiferromagnet (KLHA) model and the nature of its ground state remain unknown. Numerous candidate compounds for the spin-1/2 KLHAs have been experimentally identified [6-13]. While these materials exhibit gapless quantum spin liquid (QSL) behaviors, their microscopic models significantly deviate from the ideal KLHA model due to structural disorders caused by site-mixing and/or anisotropic Dzyaloshinsky–Moriya (DM) interactions [14,15].

Among the diverse family of kagome systems, spin-1 KLHAs stand out. Governed by both quantum and thermal fluctuations, these compounds have been theoretically predicted to host a variety of exotic ground states, such as trimer phases, hexamer phases, and QSL [16-18]. The richness of the predicted magnetic behaviors underscores the fundamental importance of these systems in the broader context of quantum magnetism.

The exploration of spin-1 KLHAs is of considerable interest within the field of magnetism, primarily due to their distinctive magnetic properties and the relatively sparse variety of known materials [19-25]. Among the fundamental magnetic characteristics of both classical and quantum KLHAs are the 120° spin configuration and the magnetic-field-induced 1/3 magnetization plateau [26-28]. However, an experimental realization of a spin-1 KLHAs that demonstrates both properties simultaneously remains elusive.

This study aims to develop new materials that can act as exemplary model systems for spin-1 KLHAs, thus facilitating a deeper understanding of their theoretical and practical implications in magnetism. In this context, we introduce the newly synthesized compound BaNi$_3$(AsO$_4$)$_2$(OH)$_2$ with isostructure of kagome materials like as β-vesignieite BaCu$_3$(VO$_4$)$_2$(OH)$_2$ [29], BaCo$_3$(VO$_4$)$_2$(OH)$_2$ [30], and β-BaNi$_3$(VO$_4$)$_2$(OH)$_2$ [31] (space group: $R\bar{3}m$), which promises to provide a new window into the intricate world of kagome antiferromagnets.

We successfully synthesized high-quality polycrystalline samples of BaNi$_3$(AsO$_4$)$_2$(OH)$_2$ using a hydrothermal method. Structural analysis via powder x-ray diffraction confirms a trigonal structure with high crystallinity and a single crystallographic Ni site, indicative of a perfect kagome lattice. Magnetic susceptibility measurements reveal predominantly ferromagnetic interactions with a Weiss temperature θ$_W$ ~ 28.5 K, while specific heat measurements and low-temperature magnetic susceptibility data indicate antiferromagnetic ordering at $T_N$ ~ 5.8 K. Additionally, isothermal magnetization measurements below $T_N$ reveal a 1/3 magnetization plateau, marking the first such observation in a spin-1 Ni$^{2+}$ kagome system. Despite the prevalent ferromagnetic interactions, the emergence of the one-third magnetization plateau underscores

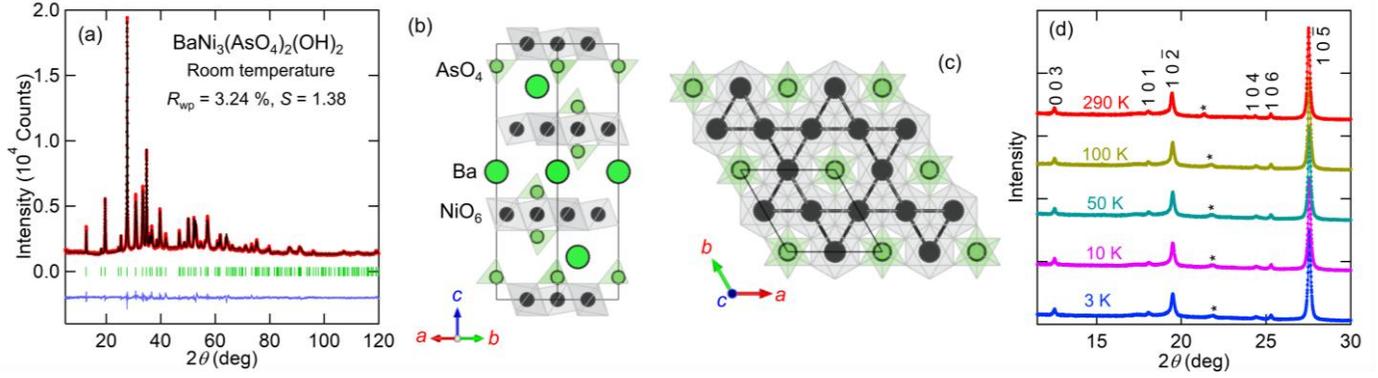

**Fig. 1** (a) Powder x-ray diffraction patterns of BaNi$_3$(AsO$_4$)$_2$(OH)$_2$. The observed intensities (red), calculated intensities (black), and their differences (blue) are shown. (b) The crystal structure of BaNi$_3$(AsO$_4$)$_2$(OH)$_2$ viewed along the $c$ axis. (c) The kagome layer viewed along the $ab$ plane. (d) Low-temperature X-ray diffraction (XRD) profiles were measured from room temperature down to 3 K.

its remarkable robustness and stability in Kagome magnets. Furthermore, application of the Clausius–Clapeyron theorem potentially indicates that the accompanying antiferromagnetic ground state arises from the A-type arrangement of in-plane ferrimagnetic order in the one-third plateau phase. This insight not only highlights the intrinsic resilience of the one-third plateau but also provides a deeper understanding of the interplay between ferrimagnetic and antiferromagnetic configurations in Kagome lattice systems.

## II. Experimental Methods

Polycrystalline sample of BaNi$_3$(AsO$_4$)$_2$(OH)$_2$ were prepared using the hydrothermal method, modified from the synthesis method used for BaCo$_3$(VO$_4$)$_2$(OH)$_2$ [30]. All starting materials were purchased from FUJIFILM Wako Pure Chemical Corporation. A Teflon beaker of 30 ml volume containing 0.20 g of Ba(OH)$_2$·8H$_2$O, 0.28 g of Ni(NO$_3$)$_2$·6H$_2$O, 0.16 g of Na$_2$HAsO$_4$, and 10 ml of pure H$_2$O, was heated at 170°C for 60 h. In the synthesis, turquoise-green-colored powders were obtained after rinsing with distilled water several times and drying at 110°C.

The thus-obtained samples were characterized by laboratory powder x-ray diffraction XRD (MiniFlex600, Rigaku) using Cu-Kα radiation. The cell parameters and crystal structure were refined by the Rietveld method using the Z-RIETVELD software [32]. Low-temperature XRD (SmartLab, Rigaku) was measured with Cu-Kα in the temperature range of 3–290 K. The temperature dependence of magnetization was measured under magnetic fields up to 7 T in a magnetic property measurement system (MPMS; Quantum Design). The temperature dependence of heat capacity was measured by a conventional relaxation method in a physical property measurement system (PPMS; Quantum Design).

**TABLE I** Crystallographic parameters for BaNi$_3$(AsO$_4$)$_2$(OH)$_2$ (space group: $R\bar{3}m$) determined from powder x-ray diffraction experiments. The obtained lattice parameters are $a = 5.806\,62(4)$ Å, $c = 21.0194(2)$ Å. $B$ is the atomic displacement parameter. The column labeled "Occ." indicates the site occupancy for each atomic position.

| atom | site | Occ. | $x$ | $y$ | $z$ | $B$ (Å$^2$) |
|---|---|---|---|---|---|---|
| Ba1 | 3$b$ | 1 | 2/3 | 1/3 | 5/6 | 0.16(2) |
| Ni1 | 9$e$ | 1 | 5/6 | 2/3 | 2/3 | 0.98(4) |
| As1 | 6$c$ | 1 | 1/3 | =1−x | 0.75302(4) | 0.41(8) |
| O1 | 6$c$ | 1 | 1/3 | =1−x | 0.8342(2) | 0.84(3) |
| O2 | 18$h$ | 1 | 0.4978(3) | =1−x | 0.72760(8) | 0.55(6) |
| O3 | 6$c$ | 1 | 0 | 0 | 0.7161(2) | 0.19(3) |

Magnetization curves up to approximately 50 T were measured by the induction method in a multilayer pulsed magnet at the International Mega Gauss Science Laboratory in the Institute for Solid State Physics.

## III. Results

Figure 1 show the powder XRD pattern of BaNi$_3$(AsO$_4$)$_2$(OH)$_2$. Every diffraction peak can be indexed on the basis of a trigonal structure of space group of $R\bar{3}m$ with a lattice consisting of $a = 5.806\,62(4)$ Å, $c = 21.0194(2)$ Å. The observed narrowness of the peak width signifies an exemplary crystallinity inherent in the sample. Utilizing the structural parameters of isostructural kagome materials [29-31] as an initial structural model, we refined the atomic positions by the Rietveld refinement method. The powder pattern is perfectly reproduced by the crystal structure with the atomic coordinates delineated in Table I; contributions from hydrogen atoms were ignored in the structural refinement. Following the approach previously employed to identify potential low-symmetry forms in BaNi$_3$(VO$_4$)$_2$(OH)$_2$ [31], we also tested a $C2/m$ model for BaNi$_3$(AsO$_4$)$_2$(OH)$_2$. However, both chemical arguments and crystallographic refinements strongly support the $R\bar{3}m$ solution as the most accurate description. Details of all refinements, unit-cell transformations, and the MEM analyses are provided in the Supporting Materials [33].

To further verify the stability of the $R\bar{3}m$ phase, we

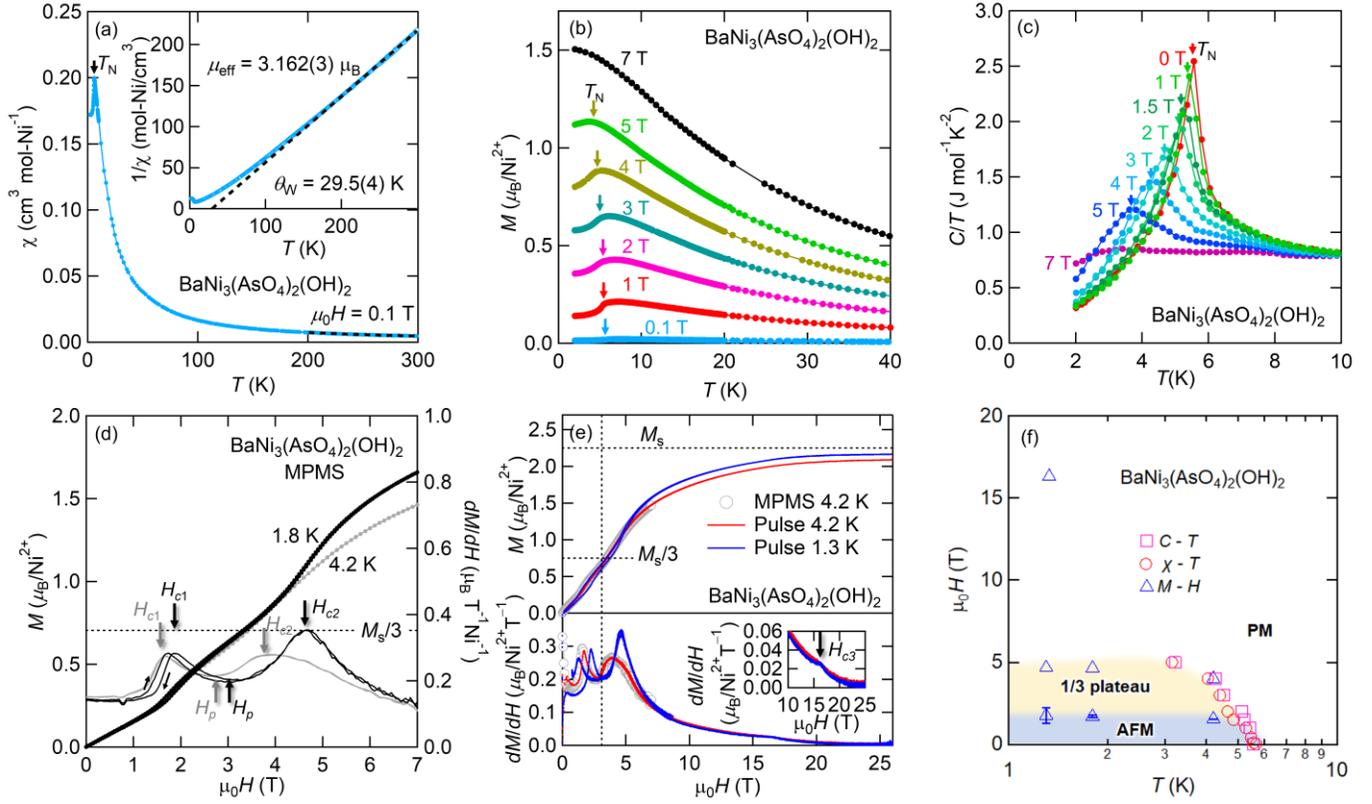

**Fig. 2** (a) Temperature dependence of the magnetic susceptibility $\chi = M/H$ (main panel) and inverse magnetic susceptibility $1/\chi$ (inset) for $BaNi_3(AsO_4)_2(OH)_2$. The dotted line on the $1/\chi$ data represents a fit to the Curie-Weiss model. (b) Temperature dependences of $M$ measured for sereval magnetic fields from 1 T to 7 T. The arrows denote themperature signifying magnetic anomalies, and their colors aligns with those of the data under the corresponding applied magnetic field. (c) The heat capacity divided by temperature $C/T$ of $BaNi_3(AsO_4)_2(OH)_2$ under various magnetic fields from 0 T to 7 T. (d)The isothermal magnetization curves $M$ and their derivative $dM/dH$ measured at 1.8 K and 4.2 K for $BaNi_3(AsO_4)_2(OH)_2$. The horizontal dashed line corresponds to the one-third magnetization value $M_s/3$. (e) The $M$ data at 4.2 K under pulsed magnetic fields up to 26 T measured at 1.3 K and 4.2 K, indicating the saturation magnetization $M_s = gS = 2.25\ \mu_B$. The vertical dashed line indicates the local minimum of the $dM/dH$ curve, suggesting the presence of magnetization plateau. The horizontal dashed line in the main panel corresponds to the one-third magnetization value $M_s/3$. (f) Magnetic field vs temperature phase diagram for $BaNi_3(AsO_4)_2(OH)_2$ constructed by magnetic measurement, heat capacity measurement, and high-field magnetization data. Squares, circles, and triangles are the transition temperatures and fields determined from the temperature dependence of heat capacities, magnetic susceptibilities, and isothermal magnetization curves, respectively. Bars represent the hysteresis field ranges of first-order phase transitions.

performed low-temperature X-ray diffraction measurements from room temperature down to 3 K, as shown in Fig. 1(d). Although a structural transition from $R\bar{3}m$ to $C2/m$ would typically manifest as splitting of the $h \neq k$ Bragg reflections (e.g., 101, $10\bar{2}$, and $10\bar{5}$), we observed no splitting at any temperature within this range. Additional temperature-dependent XRD data over the entire examined angular range are available in the Supplemental Materials [33], providing compelling evidence that no splitting occurs in any $h \neq k$ Bragg reflection over the entire temperature range. From these results, we concluded that only one crystallographic Ni site exists in the $ab$-plane, characterized by a single Ni-Ni distance. This observation underscores the realization of a perfect spin-1 kagome lattice.

Figure 2(a) shows the temperature-dependent magnetic susceptibility ($\chi$) and its inverse for $BaNi_3(AsO_4)_2(OH)_2$ measured at $\mu_0H = 0.1$ T. A Curie-Weiss fitting of the inverse susceptibility in the 200–300 K range yields an effective magnetic moment of $\mu_{eff} = 3.181(3)\ \mu_B$ and a Weiss temperature of $\theta_W = 28.5(5)$ K. For the title compound, a high $g$-value of $g = 2.25$ has been estimated for $Ni^{2+}$ ($S = 1$) spins. This large $g$-value can be explained through the concept of pseudo-quadrupole and anisotropic exchange interactions involving excited states of $Ni^{2+}$ ions, as theoretically discussed by Yosida et al [34]. The obtained effective moment is typical for $Ni^{2+}$ spins. The positive $\theta_W$-value indicates predominantly ferromagnetic interactions among the $S = 1$ spins.

At lower temperatures, a single peak is observed, indicating antiferromagnetic order. The Néel temperature $T_N$ is estimated to be 5.8 K from the peak position. Although the $\theta_W$-value is positive, the manifestation of antiferromagnetic ordering at $T_N$ < $\theta_W$ signifies the concurrent presence of both ferromagnetic and antiferromagnetic interactions. This phenomenon is attributable to the pronounced ferromagnetic nature of the higher-order nearest-neighbor interactions within the kagome lattice structure. A comprehensive analysis of the validity of this spin model will be discussed later.

Figure 2(b) expands the low-temperature region of $\chi$ measured under magnetic fields ranging from 1 to 7 T. As the applied magnetic field increases, the magnetic ordering temperature shifts to lower values, and the susceptibility drop associated with magnetic ordering becomes less pronounced. At 7 T, the magnetic ordering is suppressed below 2 K.

Evidence for magnetic ordering in $BaNi_3(AsO_4)_2(OH)_2$ is demonstrated by both specific heat measurements and magnetic susceptibility data. Heat capacity measurements reveal a $\lambda$-type peak at approximately 5.8 K, characteristic of an antiferromagnetic transition, when no magnetic field is applied. This peak, shown in Fig. 2(c), corresponds well with the magnetic anomaly observed in the low-field $\chi$-data (see Fig. 2(b)). Additionally, increasing the magnetic field causes this peak to shift to lower temperatures, consistent with the changes seen in the $\chi$-data. Notably, at an applied field of 7 T, the peak disappears entirely within the measured temperature range. This alignment between the heat capacity and susceptibility measurements strongly supports the presence of bulk antiferromagnetic ordering in $BaNi_3(AsO_4)_2(OH)_2$.

To elucidate the magnetic ordering and interactions within $BaNi_3(AsO_4)_2(OH)_2$, we conducted isothermal magnetization measurements below $T_N$, as illustrated in Fig. 2(d). At lower magnetic fields, the $M$-$H$ curves at 1.8 K and 4.2 K demonstrate a two-step increase, marked by distinct peaks in the $dM/dH$ plots. Specifically, at 4.2 K, the $dM/dH$ peak occurs at $\mu_0 H_{c1}$ = 1.55 T, whereas at 1.8 K, during the field up-sweep, this peak shifts to $\mu_0 H_{c1}$ = 1.85 T, and during the down-sweep, it decreases to $\mu_0 H_{c1}$ = 1.60 T. This behavior indicates pronounced hysteresis and suggests a first-order field-induced phase transition at $H_{c1}$. Furthermore, at 1.8 K, the high-field phase transition is marked by a $dM/dH$ peak at $\mu_0 H_{c2}$ = 4.6 T, whereas at 4.2 K, this peak not only broadens but also shifts to a lower field, $\mu_0 H_{c2}$ = 3.85 T. These isothermal magnetization processes reveal that the saturation magnetization of $Ni^{2+}$ spins is not achieved at either temperature.

To delve deeper into the magnetization process, we performed pulsed-field magnetization measurements up to 26 T, as shown in Fig. 2(e). At these high fields, the magnetization approaches an asymptotic value of approximately $M_s$. Notably, at 1.3 K, these pulsed-field measurements indicate increased hysteresis at $\mu_0 H_{c1}$ compared to the steady-field measurements at 1.8 K. Additionally, at 1.3 K, a slight jump in magnetization near the saturation field, at $\mu_0 H_{c3}$ = 16.5 T, was observed, as detailed in the inset of Fig. 2(e).

Considering thermal effects, the dips in the $dM/dH$ curves, formed by peaks at $\mu_0 H_{c1}$ and $\mu_0 H_{c2}$, align with the center of a magnetization plateau. The magnetization at this plateau, $\mu_0 H_p$, approaches the fractional value of $M/M_s$ = 1/3, confirming the presence of a 1/3 plateau. Building on these findings, magnetic susceptibility, magnetization, and heat capacity data have been systematically integrated to construct $\mu_0 H$–$T$ phase diagrams, as shown in Fig. 2(f). In cases where the transition displays hysteresis, we use the midpoint of the hysteresis loop to define the transition field, and the vertical bars represent the width of the hysteresis field range.

## IV. Discussion

Our investigations of the compound $BaNi_3(AsO_4)_2(OH)_2$ have revealed an intriguing combination of magnetic properties. Although the Curie–Weiss analysis suggests a positive Weiss temperature—indicative of dominant ferromagnetic interactions at higher temperatures—the material ultimately develops antiferromagnetic (AFM) order at low temperatures. Additionally, a metamagnetic transition emerges, leading to a magnetization plateau at approximately one-third of the full saturation value.

A noteworthy feature of the magnetic phase diagram is that the AFM–PM (paramagnetic) and 1/3–PM transition lines merge continuously, displaying nearly identical slopes. According to the Clausius–Clapeyron relation, $dH/dT = -\Delta S/\Delta M$, this implies that the net magnetization change from the AFM ground state to the PM state is effectively the same as that from the 1/3 plateau to the PM state. Consequently, the AFM ground state itself must carry a substantial net moment similar in magnitude to that observed in the 1/3 plateau.

This observation challenges the canonical view of classical kagome antiferromagnets, where a 120° spin arrangement in zero field is normally replaced by an up-up-down (UUD) configuration in the 1/3 plateau regime. In the conventional scenario, the 120° arrangement has zero net moment, while the UUD phase possesses a net magnetization equal to one-third of the saturation value. By contrast, the Clausius–Clapeyron analysis here strongly suggests that the ground state of $BaNi_3(AsO_4)_2(OH)_2$ is not a simple 120° spin structure and instead must carry a finite net magnetization. A natural possibility is that each kagome layer exhibits a ferromagnetic component (consistent with the positive Weiss temperature), but interlayer coupling is antiferromagnetic, thus partially canceling the overall magnetization yet leaving a residual net moment.

Bond frustration provides a powerful framework for understanding how ferromagnetic (FM) and antiferromagnetic (AFM) interactions can coexist within a single material [35,36]. This phenomenon arises from multiple superexchange pathways that carry competing signs, leading to a delicate balance of opposing magnetic forces. Such behavior has been widely studied in chromium spinels—including $ZnCr_2S_4$, $ZnCr_2Se_4$, and $HgCr_2S_4$—which exhibit a positive Weiss temperature but ultimately adopt an AFM order distinct from the simple A-type AFM that would emerge from purely in-plane FM interactions [37–41]. Notably, the transition from a predominantly ferromagnetic character at higher temperatures to an antiferromagnetic ground state at lower temperatures is also found in other frustrated lattices, such as $NiBr_2$ [42-44], $FeI_2$ [45], $AgCrSe_2$ [46] on triangular lattices, and $Cu_3Y(SeO_3)_2O_2Cl$ [47] on a distorted kagoem lattice. In these systems, frustration is further enhanced by the presence of closed loops that host conflicting interactions,

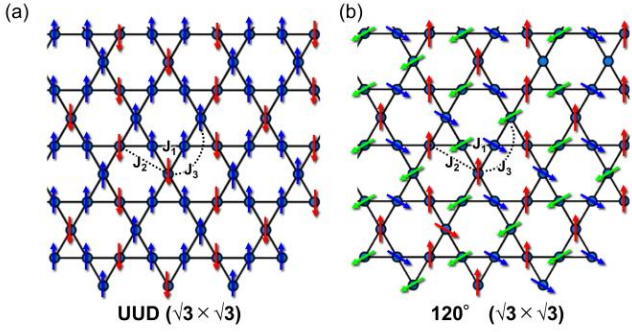

**Fig. 3** Two possible ground-state spin configurations of the kagome-lattice antiferromagnet with ferromagnetic $J_2$ couplings at wave vector $\sqrt{3} \times \sqrt{3}$ reflecting a distinct symmetry that can be probed through various magnetic measurements. (a) A collinear "up-up-down" (UUD) arrangement forming an A-type antiferromagnetic stacking sequence, in which the net moment reverses between successive layers. (b) A coplanar 120° order with $\sqrt{3} \times \sqrt{3}$ structure.

generating a multitude of nearly degenerate spin configurations.

In a similar vein, the inherently frustrated geometry of the kagome lattice in $BaNi_3(AsO_4)_2(OH)_2$ may explain both its positive Weiss temperature at high temperatures and the emergence of more complex magnetic states—including the appearance of a 1/3 magnetization plateau—at lower temperatures. The intricate interplay of multiple competing superexchange pathways within the kagome lattice thus provides a compelling example of how bond frustration can lead to rich and unconventional magnetic behavior in low-dimensional systems.

From a mean-field perspective, both the Weiss temperature and the saturation field can be approximated by the exchange couplings $J_1$, $J_2$, and $J_3$. Consequently, the ratio of $-\theta_W/\mu_0 H_{sat}$ depends only on the ratios $J_2/|J_1|$ and $J_3/|J_1|$. Because these constraints are relatively weak, an infinite number of numerical solutions exists. However, focusing on the solution that minimizes the absolute value of $J_2/|J_1|$ gives $J_2/|J_1| \sim 1.72$ and $J_3/|J_1| \sim 0.42$. This representative solution demonstrates that, while $J_2$ locally promotes ferromagnetic alignment, the overall magnetic order is not simply ferromagnetic. As theorized by Messio et al. [48], ferromagnetism is stabilized when $-J_2/|J_1| < -1$ and $-J_3/|J_1| < -1$. In contrast, for $-J_3/|J_1| > -1$, even strong ferromagnetism $J_2$ interactions leads to $\sqrt{3} \times \sqrt{3}$ 120° magnetic order. Our numerical calculations broadly corroborate the theoretical findings of Messio et al [48]. Although the complete set of interactions produces a Weiss temperature of 29.5 K—signifying a considerable net ferromagnetic tendency—the actual ground state remains non-ferromagnetic (see Supplemental Materials [33] for further details).

In the $\sqrt{3} \times \sqrt{3}$ UUD or 120° configuration, spins within each crystallographic unit cell are arranged so that those connected by $J_2$ remain parallel, while any frustration associated with $J_1$ is still accommodated. For instance, in the simpler $\sqrt{3} \times \sqrt{3}$ UUD arrangement, two spins point "up" and one points "down," repeating this motif throughout the lattice and thereby preserving parallel alignment on $J_2$ bonds. Likewise, in the 120° phase, ferromagnetic alignment persists along $J_2$ links, allowing strong local ferromagnetic interactions without forcing the entire system into a fully ferromagnetic state. As illustrated in Figure 3, these configurations effectively capitalize on the strong $J_2$ coupling yet avoid the energetic cost of a uniform magnetic order. Moreover, these spin arrangements show that the strong ferromagnetic $J_2$ interaction itself further stabilizes these specific configurations. Recent calculations employing the orthogonalized finite-temperature Lanczos method, as reported by Morita, provide compelling numerical evidence that the ferromagnetic $J_2$ interaction in the kagome lattice is pivotal role for stabilizing the characteristic $\sqrt{3} \times \sqrt{3}$ UUD spin configuration [49]. Consequently, although the overall network of interactions leads to a positive Weiss temperature, the actual low-temperature spin ordering need not be purely ferromagnetic. Based on these numerical calculations and discussions, these findings strongly suggest that a viewpoint grounded in bond frustration provides a robust framework for understanding the fundamental physics of $BaNi_3(AsO_4)_2(OH)_2$.

Bond frustration naturally fosters intermediate-field phases below the threshold for complete magnetic saturation. For instance, $HgCr_2S_4$ [41], $NiBr_2$ [42], $FeI_2$ [45], and $AgCrSe_2$ [46] are all bond-frustrated magnets that exhibit positive Weiss temperatures and show an "intermediate" phase before transitioning into a forced ferromagnetic state at higher fields. Although these examples differ from $BaNi_3(AsO_4)_2(OH)_2$ in their specific lattice geometries, they share three critical features: a frustrated magnetic network, a positive Weiss temperature reflective of underlying ferromagnetic tendencies, and partial spin alignment in intermediate fields driven by competing interactions. By analogy, in $BaNi_3(AsO_4)_2(OH)_2$, the metamagnetic transition resulting in the 1/3 magnetization plateau can be understood as a similar partial spin alignment across particular layers or clusters, rather than a straightforward evolution from a conventional 120° arrangement with zero net moment. This perspective unifies various frustrated systems in which multiple exchange pathways give rise to pronounced competition, culminating in intermediate field-driven states.

Several lines of evidence support the notion that bond frustration and anisotropy govern the magnetic behavior of $BaNi_3(AsO_4)_2(OH)_2$. One key observation is that Curie–Weiss analyses yield an effective magnetic moment larger than the spin-only value for $Ni^{2+}$ (with $g \approx 2$), suggesting an enhanced $g$-factor of about 2.25. This elevation in g reveals significant single-ion anisotropy or anisotropic exchange, thereby indicating that a simple isotropic Heisenberg model cannot fully describe the system. Additionally, magnetization measurements confirm the presence of a 1/3 plateau, reinforcing the idea that a partially ordered phase emerges under an applied field.

A particularly revealing comparison can be drawn with $CoCl_2 \cdot 2H_2O$ [50], which displays strong single-ion anisotropy and a well-characterized one-third magnetization plateau.

Likewise, vanadium-based kagome fluorides demonstrate that pronounced Ising anisotropy can stabilize a 1/3 plateau on a kagome lattice [51]. Furthermore, the Ising triangular lattice in $FeI_2$ exhibits multiple magnetization plateaus [45]. Collectively, these examples underscore the pivotal role of anisotropy in inducing partial spin polarization within frustrated magnetic lattices, illustrating how directional or single-ion effects can favor discrete plateaus that would otherwise be suppressed in fully isotropic systems.

By analogy, $BaNi_3(AsO_4)_2(OH)_2$—which also crystallizes in a kagome lattice—appears to adopt an antiferromagnetic ground state that can be described by a collinear UUD spin configuration or a closely related arrangement. In such a state, partial spin alignment generates a net moment equal to one-third of the fully saturated value, consistent with the 1/3 plateau observed in the $H$–$T$ phase diagram. Although the actual spin structure may be more complex than a simple UUD model—potentially involving weak ferromagnetic components arising from non-collinear orders such as helical arrangements—any configuration that yields a one-third net moment aligns with the experimental findings. Overall, these observations demonstrate that the interplay of bond frustration, anisotropy, and specific collinear spin configurations provides a coherent framework for explaining the unconventional 1/3 magnetization plateau in $BaNi_3(AsO_4)_2(OH)_2$.

One might expect that once the 1/3 magnetization plateau is reached under an external field of around 3 T, the magnetization would either remain at this plateau or even grow stronger as the temperature is lowered. Surprisingly, however, the experimental data reveal a pronounced peak in the magnetization near 6 K, followed by a decrease upon further cooling. This unexpected trend can be understood by noting that short-range ferromagnetic correlations become dominant just above the Néel temperature (approximately 5.8 K). As thermal fluctuations subside, these correlations temporarily boost the overall magnetization. Once the temperature drops below the Néel point, the system settles into a long-range antiferromagnetic state. Although the 1/3 plateau persists in principle, the antiferromagnetic ordering partially cancels out the net magnetic moment, ultimately reducing the observed magnetization. This competition between ferromagnetic short-range order above the Néel temperature and long-range antiferromagnetic ordering below it offers a coherent explanation for the peak-and-drop behavior of the magnetization under these conditions.

Hence, the interplay between strong ferromagnetic correlations at higher temperatures and the establishment of a global antiferromagnetic order at lower temperatures underlies the antiferromagnetic magnetization-drop. Similar phenomena occur in other frustrated materials that exhibit local ferromagnetic tendencies yet ultimately settle into an antiferromagnetic ground state [41-44], resulting in unconventional temperature dependence of the magnetization.

Looking ahead, direct measurements of the spin structure under external magnetic fields—such as neutron diffraction, polarized neutron scattering, or x-ray magnetic scattering—are essential for clarifying the real-space spin arrangement in both the AFM ground state and the 1/3-plateau phase. These techniques would establish whether each kagome layer carries a net moment that alternates sign layer by layer or whether more intricate cluster-based spin patterns are involved. In addition, local-probe experiments such as nuclear magnetic resonance (NMR) or muon spin rotation (μSR) would help confirm the presence of distinct internal fields associated with these competing states, offering deeper insight into the fundamental role of bond frustration in Ni-based kagome magnets.

## V.  Summary

We have successfully synthesized $BaNi_3(AsO_4)_2(OH)_2$ via hydrothermal synthesis, confirming it as an ideal model material for spin-1 KLHA. Powder XRD analysis verifies a trigonal structure with high crystallinity and a single crystallographic Ni site, indicative of an ideal kagome lattice. Magnetic susceptibility measurements reveal predominantly ferromagnetic interactions, with a Weiss temperature of ~28.5 K and an effective magnetic moment typical for $Ni^{2+}$ spins. Despite these tendencies, antiferromagnetic ordering occurs at a Néel temperature $T_N$ ~ 5.8 K, suggesting a coexistence of ferromagnetic and antiferromagnetic interactions. Isothermal magnetization and pulsed-field measurements identify first-order field-induced phase transitions and a 1/3 magnetization plateau. The nearly identical slopes of the continuous AFM–PM and 1/3-plateau–PM phase boundaries, combined with a positive Weiss temperature, strongly indicate that the ground state maintains a ferrimagnetic net moment comparable to that of the 1/3 plateau. This result departs from the conventional 120°-to-UUD picture often assumed for kagome magnets, suggesting instead that multiple, potentially bond-frustrated exchange pathways stabilize in-plane ferrimagnetic correlations. These correlations could manifest as a collinear UUD configuration or more complex arrangements, emerging in both the zero-field ground state and the field-induced metamagnetic phases. Consequently, these findings offer new directions for exploring the interplay between geometric and bond frustration in kagome and related lattice geometries, highlighting the need for further experimental and theoretical work to fully clarify the underlying spin configurations.

### Acknowledgement

This work was supported by JST PRESTO Grant Number JPMJPR23Q8 (Creation of Future Materials by Expanding Materials Exploration Space) and JSPS KAKENHI Grant Numbers. JP23H04616 (Transformative Research Areas (A) "Supra-ceramics"), JP24H01613 (Transformative Research Areas (A) "1000-Tesla Chemical Catastrophe"), JP22K14002 (Young Scientific Research), and JP24K06953 (Scientific Research (C)). Part of this work was carried out by joint research in the Institute for Solid State Physics, the University of Tokyo (Project Numbers 202211-GNBXX-0034, 202211-GNBXX-0035, 202211-HMBXX-0022, 202306-GNBXX-0128, 202306-GNBXX-0129, and 202306-HMBXX-0072).